\documentclass[aps,prc,twocolumn,groupedaddress,showpacs]{revtex4}

\usepackage{graphicx}

\begin{document}

\title{Density dependence of the symmetry energy and the nuclear equation of state: A Dynamical and
Statistical model perspective}
\author{D.V. Shetty}
\author{S.J. Yennello}
\author{G.A. Souliotis}
\affiliation{Cyclotron Institute, Texas A$\&$M University, College Station, TX 77843, USA}
\date{\today}

\begin{abstract}
The density dependence of the symmetry energy in the equation of state of isospin asymmetric nuclear 
matter is of significant importance for studying the structure of systems as diverse as the neutron-rich nuclei 
and the neutron stars. A number of reactions using the dynamical and the statistical models of 
multifragmentation, and the experimental isoscaling observable, is studied to extract information on the density dependence of 
the symmetry energy. It is observed that the dynamical and the statistical model calculations give
consistent results assuming the sequential decay effect in dynamical model to be  small.
A comparison with several other independent studies is also made to obtain important constraint 
on the form of the density dependence of the symmetry energy. The comparison rules out an extremely `` stiff " and 
`` soft " form of the density dependence of the symmetry energy with important implications for astrophysical and 
nuclear physics studies.
\end{abstract}

\maketitle

\section{Introduction}
The fundamental goal of nuclear physics is to understand the basic building blocks of nature - neutrons 
and protons - and the nature of interaction that binds them together into nuclear matter. Studying the  
nature of matter and the strength of nuclear  interaction is key to understanding some of the fundamental 
problems such as, How are elements formed? How do stars explode into supernova? What kind of matter exists 
inside a neutron star? How are neutrons compressed inside a neutron star to density trillions of times 
greater than on earth ? What determines the density-pressure relation, the so-called equation of state? 
\par
The key ingredient for constructing the nuclear equation of state is the basic nucleon-nucleon interaction.
Until now our understanding of the nucleon-nucleon interaction has come from studying nuclear matter  that 
is  symmetric in isospin (neutron-to-proton ratio, $N/Z$ $\approx$ 1) and matter found near 
normal nuclear density ($\rho_{o}$ $\approx$ 0.16 $fm^{-3}$). It is not known how far this understanding 
remains valid as one goes away from the normal nuclear density and symmetric nuclear matter. Various 
interactions used in `` ab initio " microscopic calculations predict different forms of the nuclear equation 
of state above and below the normal nuclear matter density, and away from the symmetric nuclear 
matter \cite{DIE03,WIR88,LEE98,LIU02,KAI02,FUC06}.  As a result, the symmetry energy, which is the difference 
in energy between the pure neutron matter and the symmetric nuclear matter, shows very different behavior 
above and below normal nuclear density \cite{FUC06} (see Fig. 1). 
\par
In general, two different forms of the density dependence of the symmetry energy have been predicted. One, 
where the symmetry energy increases monotonically with increasing density (`` stiff " dependence) and the 
other, where the symmetry energy increases initially up to normal nuclear density and then decreases at 
higher densities (`` soft " dependence). Constraining the form of the density dependence of the 
symmetry energy is important not only for a better understanding of the nucleon-nucleon interaction, 
and hence its extrapolation to the  structure of neutron-rich nuclei \cite{BRO00,HOR01,FUR02,OYA98}, but 
also for determining the structure of compact stellar objects such as neutron 
stars \cite{LAT91,LEE96,PET95,LAT00,HIX03,LAT04,STE05}. 
\begin{figure}
\resizebox{0.48\textwidth}{!}{
\includegraphics{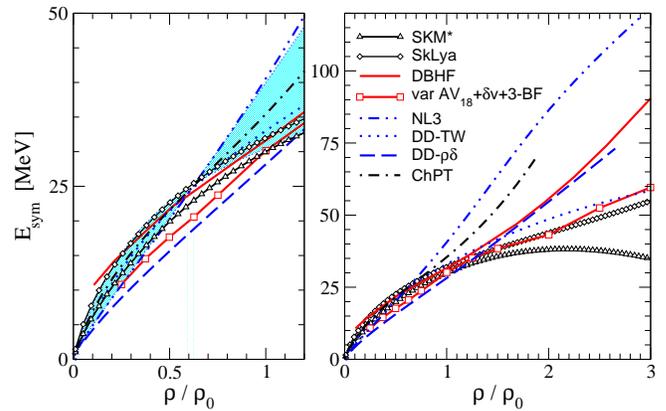}
}
\caption{(Color online) Symmetry energy as a function of density predicted by microscopic `` ab initio " 
calculations. The left panel shows the low-density  region, while the right panel displays the high-density 
range. The figure is taken from Ref. \cite{FUC06}.}
\label{fig:1}       
\end{figure}
For example, a `` stiff " form of the density dependence of the symmetry energy is predicted to lead to a 
large neutron skin thickness compared to a `` soft " dependence \cite{HOR01,OYA98,HORO01,HOR02,STO03}. 
Similarly, a `` stiff " dependence of the symmetry energy can result in rapid cooling of a neutron star, 
and a larger neutron star radius, compared to a `` soft " density dependence of the symmetry energy 
\cite{STO03,LAT94,SLA02}. The nuclear Equation Of State (EOS) is therefore a fundamental entity that 
determines the properties of systems as diverse as atomic nuclei and neutron stars, and the knowledge 
of which is of significant importance \cite{DAN02,LAT04,STE05}. 
\par
Experimentally, the best possible  means of studying the  nuclear equation  of state  at sub-normal  nuclear 
density is through intermediate-energy heavy-ion reactions \cite{BAL98,BAR05}. In this kind of reaction, 
an excited nucleus (the composite of the projectile and the target nucleus) expands to a sub-nuclear density  
and disintegrates into  various light and heavy fragments in a process called multifragmentation. By studying 
the isotopic yield  distribution  of these  fragments one  can  extract important information about the 
symmetry energy and its density dependence. Current studies on the nuclear equation of state are  
limited to beams consisting of stable nuclei. It is hoped that in the future radioactive beam facilities
such as, FAIR (GSI) \cite{GSI}, SPIRAL2 (GANIL) and FRIB (USA) \cite{RIA} will provide tremendous 
opportunities for exploring the nuclear EOS in regions never before studied ({\it {i.e.}}, extreme isospin 
and away from normal nuclear density). 
\par
In this work, we have made an attempt to study the density dependence of the symmetry energy using two 
different theoretical approaches for studying multifragmentation, namely the dynamical and 
the statistical model approaches of multifragmentation. In section II, the isoscaling technique used to
study the density dependence of the symmetry energy, and their different interpretations in terms of 
statistical and dynamical approach, are presented. In section III and IV, a brief description of the
experiment and the experimental results are presented. The dynamical and the statistical approaches used to
interpret the experimental results are presented in section V. A comparsion between the two approaches 
with other independent studies is presented in section VI. Finally, a discussion and summary, and conclusions 
are presented in section VII and VIII, respectively.

\section{Symmetry energy and the isotopic yield distribution}

It has been shown from experimental measurements that the ratio of the fragment yields, R$_{21}$($N$,$Z$), 
taken from two different multifragmentation reactions, 1 and 2,  obeys an exponential dependence on the 
neutron number ($N$) and the proton number ($Z$) of the fragments; an observation known as 
isoscaling \cite{TSANG01,XU00,SHE03}. The dependence is characterized by the relation,

\begin{equation}
       R_{21}(N,Z) = Y_{2}(N,Z)/Y_{1}(N,Z) = Ce^{(\alpha N + \beta Z)}
\end{equation}

Where, $Y_{2}$ and $Y_{1}$ are the fragment yields from the neutron-rich and the neutron-deficient systems, 
respectively. $C$ is an overall normalization factor, and $\alpha$ and $\beta$ are the parameters 
characterizing the isoscaling behavior. Isoscaling is also theoretically predicted by the dynamical
\cite{ONO03,QIN06,LIU04,COL05,DOR06} and statistical \cite{BOT02,RAD02,FRI90,GRO97} models
of multifragmentation. In these models, the difference in the chemical potential of systems with 
different neutron-to-proton ration ($N/Z$) is directly related to the isoscaling parameter $\alpha$. The
isoscaling parameter $\alpha$, is  related to the symmetry energy $C_{sym}$, through the relation, 

\begin{equation}
           \alpha = \frac{4C_{sym}}{T} \bigg (\frac{Z_{1}^{2}}{A_{1}^{2}} - \frac{Z_{2}^{2}}{A_{2}^{2}}\bigg )
\end{equation}

where, $Z_{1}$, $A_{1}$ and $Z_{2}$, $A_{2}$ are the charge and the mass numbers from the two
systems and $T$ is the temperature. This relation provides a simple and straight-forward connection between 
the symmetry energy and the fragment isotopic yield distribution.
\par
It must be mentioned that although the above equation derived from the statistical and the dynamical models 
of multifragmentation appears similar in form, the physical meaning of the terms involved in this equation
differ for each model.
\par
1) In statistical models, the $Z/A$ in Eq. (2) corresponds to the charge-to-mass ratio of the initial 
equilibrated fragmenting system. Whereas, in dynamical models, it corresponds to the charge-to-mass ratio 
of the liquid phase at a certain time ($\approx$ 300 {\it {fm/c}}) during the dynamical evolution of the 
colliding systems. 
\par
2) The interpretation of the symmetry energy $C_{sym}$, in dynamical and statistical models 
also differs significantly. The dynamical models relate the symmetry energy in the above equation to that of 
the fragmenting source. The statistical models, on the other hand, relate $C_{sym}$ to that of the fragments 
formed at freeze-out. 
\par
These conceptual differences between the statistical and the dynamical models are due to the radically 
different approaches taken in the interpretation of the multifragmentation process. The different 
interpretation has also lead to conflicting results from the use of Eq. 2, due to the different sequential 
decay effects predicted for the primary fragments by each model.
\par
The isoscaling parameter $\alpha$, in Eq. 2 corresponds to the hot primary fragments which 
undergo sequential decay into cold secondary fragments. These secondary fragments are the ones that are 
eventually detected in experiments. The experimentally determined isoscaling parameter must therefore be 
corrected for the sequential decay effect before comparing it to the theoretical models. It has been observed 
that while statistical model calculations show  no significant change in the isoscaling parameter after
sequential decay \cite{TSA01}, dynamical models give contrasting results; with some showing no
significant changes \cite{TIA06}, while others showing a change of as much as 50$\%$ \cite{ONO06}. 
\par
In this work, we adopt both theoretical approaches with their respective interpretations to study the density
dependence of the symmetry energy. In particular, we use the Antisymmetrized Molecular Dynamics (AMD) 
model \cite{ONO03,ONO04} and the Statistical Multifragmentation Model (SMM) \cite{BOT02} for this study. A 
comparison between the two can provide useful insight into the physical meaning of the above equation in the 
two models.

\section{Experiment}
\subsection{Experimental Setup}
The experiments were carried out at the Cyclotron Institute of Texas A$\&$M University (TAMU) using the
K500 Superconducting Cyclotron and the National Superconducting Cyclotron Laboratory (NSCL) at Michigan
State University (MSU). Targets of $^{58}$Fe (2.3 mg$/$cm$^{2}$) and $^{58}$Ni (1.75 mg$/$cm$^{2}$) were 
bombarded with beams of $^{40}$Ar and $^{40}$Ca at 33 and 45 MeV$/$nucleon for the TAMU measurements
\cite{JOH97}, and targets of $^{58}$Fe ($\sim$ 5 mg$/$cm$^{2}$)  and $^{58}$Ni ($\sim$ 5 mg$/$cm$^{2}$) were 
bombarded with beams of $^{40}$Ar and $^{40}$Ca at 25 and 53 MeV$/$nucleon for the NSCL measurements
\cite{YEN94}. The various combinations of target and projectile nuclei allowed for a range of $N/Z$ 
(neutron-to-proton ratio) (1.04 $-$ 1.23) of the system to be studied, while keeping the total mass constant 
($A$ = 98). In a separate experiment at TAMU, beams of $^{58}$Ni and $^{58}$Fe at 30, 40, and 47
MeV/nucleon were also bombarded on self-supporting $^{58}$Ni and $^{58}$Fe targets. 
\par
The beams in the TAMU measurements were fully stripped by allowing them to pass through a thin aluminum foil 
before being hit at the center of the target inside the TAMU 4$\pi$ neutron ball \cite{SCH95}.  Light charged 
particles ($Z$ $\leq$ 2) and intermediate mass fragments ($Z$ $>$ 2) were detected using six discrete 
telescopes placed inside the scattering chamber of the neutron ball at angles  of 10$^{\circ}$, 44$^{\circ}$, 
72$^{\circ}$, 100$^{\circ}$, 128$^{\circ}$ and 148$^{\circ}$. Each telescope consisted of a gas ionization 
chamber (IC) followed by a pair of silicon detectors (Si-Si) and a CsI scintillator detector, providing three 
distinct detector pairs (IC-Si, Si-Si, and Si-CsI) for fragment identification. The ionization chamber was of 
axial field design and was operated with CF$_{4}$ gas at a pressure of 50 Torr. The gaseous medium was 6 cm 
thick with a typical threshold of $\sim$ 0.5 MeV/nucleon for intermediate mass fragments. The silicon 
detectors had an active area of 5 cm $\times$ 5 cm and were each subdivided into four quadrants. The first 
and second silicon detectors in the stack were 0.14 mm and 1 mm thick, respectively. The dynamical energy 
range of the silicon pair was  $\sim$ 16 - 50 MeV for $^{4}$He and  $\sim$ 90 - 270 MeV for $^{12}$C. The 
CsI scintillator crystals that followed the silicon detector pair were 2.54 cm in thickness and were read out 
by photodiodes.  Good elemental ($Z$) identification was achieved for fragments that punched through the IC 
detector and stopped in the first silicon detector. Fragments measured in the Si-Si detector pair also had 
good isotopic separation. Fragments that stopped in CsI detectors showed isotopic resolution up to $Z$ = 7. 
The trigger for the data acquisition was generated by requiring  a valid hit in one of the silicon detectors.
\par
The calibration of the IC-Si detectors were carried out using the standard alpha sources and by operating the 
IC at various gas pressures. The Si-Si detectors were calibrated by measuring the energy deposited by the 
alpha particles in the thin silicon and the punch-through energies  of different isotopes in the thick 
silicon. The Si-CsI detectors were calibrated by selecting points along the different light charged isotopes 
and determining the energy deposited in the CsI crystal from the energy loss in the calibrated Si detector. 
\par
The setup for the NSCL experiment consisted of 13 silicon detector telescopes placed inside the MSU 4$\pi$
Array. Four of which were placed at 14$^{\circ}$, each of which consisted of a 100 $\mu$m thick and a 1 mm
thick silicon surface-barrier detector followed by a 20 cm thick plastic scintillator. Five telescopes were
placed at 40$^{\circ}$, in front of the most forward detectors in the main ball of the 4$\pi$ Array. They
each consisted of a 100 $\mu$m surface-barrier detector followed by a 5 mm lithium drifted silicon detector.
More details can be found in Ref. \cite{YEN94}. Good isotopic resolution was obtained as  in TAMU measurements. 

\subsection{Event Characterization} 
The event characterization of the NSCL data was accomplished by detection of nearly all the coincident
charged particles by the MSU 4$\pi$ Array. Data were acquired using two different triggers; the bulk of
which was obtained with the requirement of a valid event in one of the silicon telescopes. Additional data
were taken with a minimum bias 4$\pi$ Array trigger for normalization of the event characterization. 
The impact parameter of the event was determined by the mid-rapidity charge detected in the 4$\pi$ Array
as discussed in Ref. \cite{OGI89}. The effectiveness of the centrality cuts was tested by comparing the 
multiplicity of events from a minimum bias trigger with the multiplicity distribution when a valid fragment 
was detected at 40$^{\circ}$ \cite{YENN94}. The minimum bias trigger had a peak multiplicity of charged 
particles of one, whereas with the requirement of a fragment at 40$^{\circ}$, the peak of the multiplicity 
distribution increased to five.  
\par
The event characterization for the TAMU data was accomplished by using the 4$\pi$ neutron ball that surrounded 
the detector assembly. The neutron ball consisted of eleven scintillator tanks segmented in its median plane 
and surrounding the vacuum chamber. The upper and the lower tank were 1.5 m diameter hemispheres. Nine 
wedge-shaped detectors were sandwiched between the hemispheres. All the wedges subtended 40$^{\circ}$
in the horizontal plane. The neutron ball was filled with a pseudocumene-based liquid scintillator mixed with 
0.3 $\%$ (b.w.) of Gd salt (Gd 2-ethyl hexanoate). Scintillations from thermal neutrons captured by Gd were 
detected by twenty 5-inch phototubes : five in each hemisphere, one on each of the identical 40$^{\circ}$ 
wedges and two on the forward edges. The efficiency with which the neutrons could be detected is about 83$\%$, 
as measured with a $^{252}$Cf source.
\par
The detected neutrons were used to differentiate between the central and peripheral collisions. To understand 
the effectiveness of neutron multiplicity as a centrality trigger, simulations were carried out using a hybrid 
BUU-GEMINI calculations at various impact parameters for the $^{40}$Ca + $^{58}$Fe reaction at 33 MeV/nucleon. 
The simulated neutron multiplicity distribution was compared with the experimentally measured distribution. 
The multiplicity of neutron for the impact parameter {\it {b}} = 0 collisions was found to be higher than 
the {\it {b}} = 5 collision. By gating on the 10$\%$ highest neutron multiplicity events, one could clearly 
discriminate against the peripheral events. 
\par
To determine the contributions from noncentral impact parameter collisions,  neutrons emitted in coincidence 
with fragments at 44$^{\circ}$ and 152$^{\circ}$ were calculated at {\it {b}} = 0 {\it {fm}} and {\it {b}} = 5 {\it {fm}}. 
The number of events were adjusted for geometrical cross sectional differences. A ratio was made between the 
number of events with a neutron multiplicity of at least six, calculated at {\it {b}} = 0 {\it {fm}}, and the number 
of events with the same neutron multiplicity at {\it {b}} = 5 {\it {fm}}. The ratios were observed to be 19.0 and 
11.1 at 44$^{\circ}$ and 1.3 and 2.2 at 152$^{\circ}$ for 33 and 45 MeV/nucleon respectively. At intermediate 
angles, high neutron multiplicities were observed to be outside the region in which {\it {b}} = 5 {\it {fm}} 
contributes significantly. At backward angles the collisions at {\it {b}} = 5 {\it {fm}} made a larger contribution 
to the neutron multiplicity.
\par
In addition to the neutron multiplicity distribution, the charge distribution of the fragments was also used 
to investigate the contributions from central and mid-impact parameter collisions. The {\it {b}} = 5 
collisions produced essentially no fragments with  charge greater than three in the 44$^{\circ}$ telescope.
\par
In an earlier work \cite{JOH97}, some analysis of the fragment kinetic energy and charge distributions were
presented. It was shown that at a laboratory angle of 44$^{\circ}$ the kinetic energy and the charge
distributions are well reproduced by the statistical model calculation. Using a moving source analysis of
the fragment energy spectra, it was also shown that the fragments emitted at backward angles originate
from a target-like source, while those emitted at 44$^{\circ}$ originate primarily from a composite
source. In this work, we will concentrate exclusively on data from the laboratory angle of 44$^{\circ}$, 
which corresponds to the center of mass angle $\approx$ 90$^{\circ}$, to study the symmetry energy and the 
isoscaling properties of the fragments produced. The choice of this angle enables one to select events which 
are predominantly central and undergo bulk multifragmentation. The contributions to the intermediate mass 
fragments from the projectile-like and target-like sources can therefore be assumed to be minimal.

\section{Experimental Results}

\subsection{Fragment isotopic yield distribution}
\begin{figure}
\resizebox{0.48\textwidth}{!}{
\includegraphics{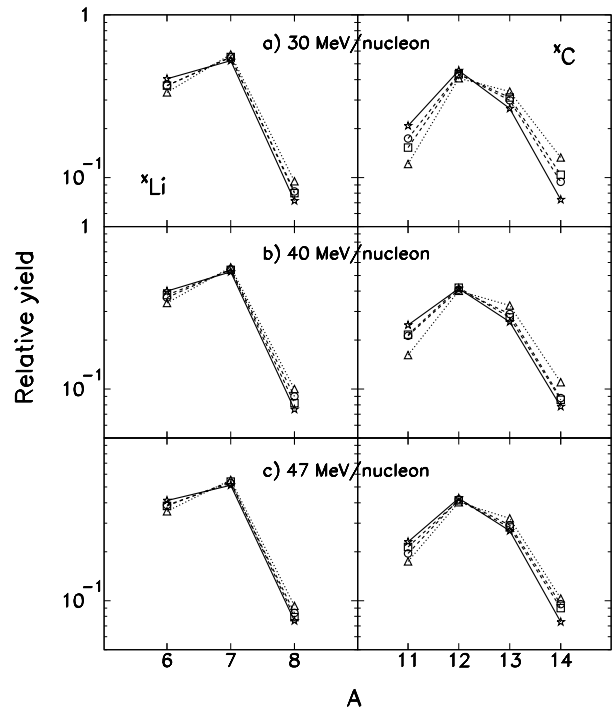}
}
\caption{Relative yield distribution of the fragments for the Lithium (left) and Carbon (right) isotopes in
$^{58}$Ni + $^{58}$Ni (stars and solid lines), $^{58}$Fe + $^{58}$Ni (circles and dashed lines), $^{58}$Ni +
$^{58}$Fe (squares and dashed lines), and $^{58}$Fe + $^{58}$Fe (triangles and dotted lines) reactions at
various beam energies.}
\label{fig:2}       
\end{figure}
\begin{figure}
\resizebox{0.48\textwidth}{!}{
\includegraphics{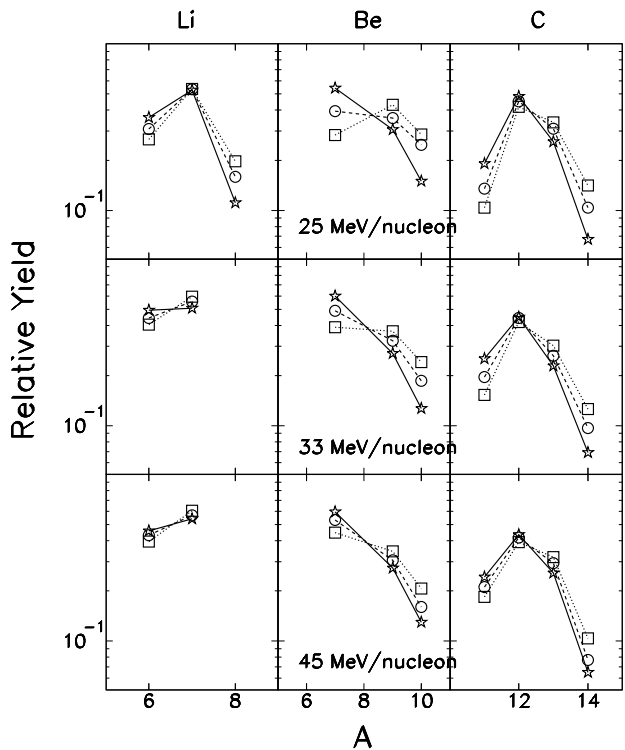}
}
\caption{Relative yield distribution of the fragments for Lithium (left), Berillium (center) and Carbon 
(right) isotopes in $^{40}$Ca + $^{58}$Ni (stars and solid lines), $^{40}$Ar + $^{58}$Ni (circles and 
dashed lines), and $^{40}$Ar + $^{58}$Fe (squares and dotted lines) reactions at various beam energies.}
\label{fig:3}       
\end{figure}
The experimentally measured relative isotopic yield distributions for the Lithium (left) and Carbon (right) 
elements, in $^{58}$Ni + $^{58}$Ni (star symbols), $^{58}$Ni + $^{58}$Fe (square symbols), $^{58}$Fe + $^{58}$Ni 
(circle symbols) and $^{58}$Fe + $^{58}$Fe (triangle symbols) reactions, are shown in Fig. 2 for beam 
energies of 30, 40 and 47 MeV/nucleon. Similarly, the isotopic yield distributions for Lithium (left), 
Berillium (center) and Carbon (right) elements, in $^{40}$Ca + $^{58}$Ni (star symbols), $^{40}$Ar + $^{58}$Ni 
(circle symbols) and $^{40}$Ar + $^{58}$Fe (square symbols) reactions, are shown in figure 3 for beam 
energies of 25, 33 and 45 MeV/nucleon. The isotope distribution for each element in Fig. 3 shows higher 
fragment yield for the neutron rich isotopes in $^{40}$Ar + $^{58}$Fe reaction (squares), which has the 
largest neutron-to-proton ratio ($N/Z$),  in comparison to the $^{40}$Ca + $^{58}$Ni reaction (stars), which 
has the smallest neutron-to-proton ratio. The yields for the reaction, $^{40}$Ar + $^{58}$Fe (circles), 
which has an intermediate value of the neutron-to-proton ratio, are in between those of the other two 
reactions. A similar feature is also observed for the $^{58}$Ni + $^{58}$Fe, $^{58}$Fe + $^{58}$Ni and 
$^{58}$Fe + $^{58}$Fe reactions shown in Fig. 2. The fragment yield distributions therefore show the 
isospin dependence of the composite system on the fragments produced in the multifragmentation reaction. One 
also observes that the relative difference in the yield distribution between the three reactions in each 
figure decreases with increasing beam energy. This is due to the secondary de-excitation of the primary 
fragments, a process that becomes important for systems with increasing neutron-to-proton ratio and 
excitation energy.

\subsection{Isotopic and Isotonic scaling}

As discussed in section II, the ratio of isotope yields in two different systems, 1
and 2, $R_{21}(N,Z) = Y_{2}(N,Z)/Y_{1}(N,Z)$, follows an exponential dependence on the
neutron number $(N)$ and the proton number $(Z)$ of the isotopes in relation known as isoscaling. 
\begin{figure}
\resizebox{0.48\textwidth}{0.5\textheight}{
\includegraphics{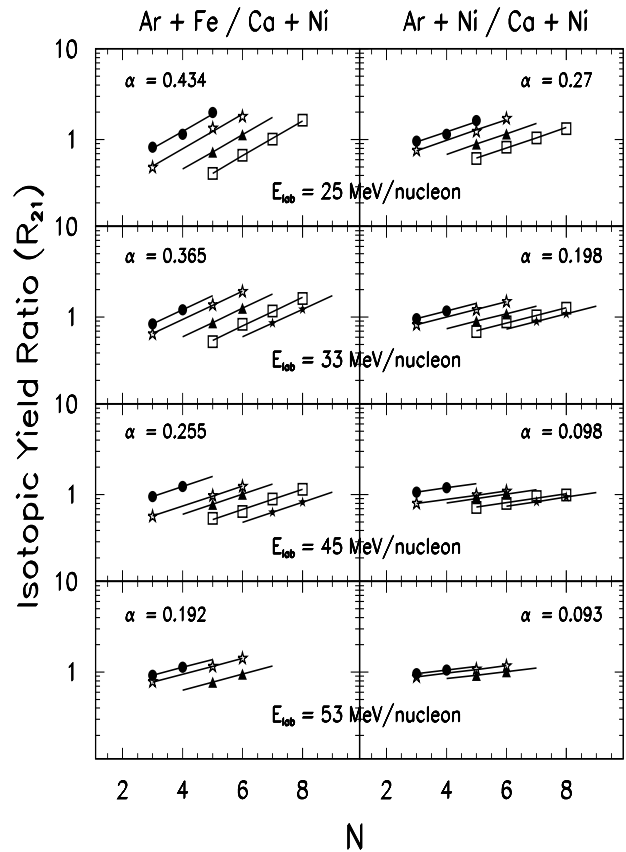}
}
\caption{Experimental isotopic yield ratios of the fragments as a function of neutron number $N$, for various
    beam energies. The left column correspond to $^{40}$Ar + $^{58}$Fe and $^{40}$Ca + $^{58}$Ni pair of 
    reactions. The right column correspond to $^{40}$Ar + $^{58}$Ni and $^{40}$Ca + $^{58}$Ni pair of 
    reactions. The different symbols correspond to $Z$ = 3 (circles), $Z$ = 4 (open stars), $Z$ = 5 
    (triangles), $Z$ = 6 (squares) and $Z$ = 7 (filled stars) elements. The lines are the exponential fits to 
    the data as explained in the text.}
\label{fig:4}       
\end{figure}
\begin{figure}
\resizebox{0.48\textwidth}{!}{
\includegraphics{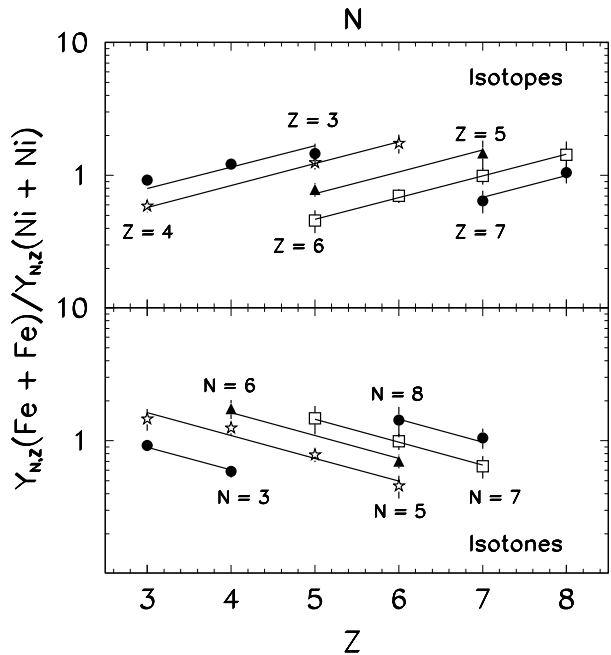}
}
\caption{Experimental isotope yield ratios (top) and isotone yield ratios (bottom) from $^{58}$Fe + $^{58}$Fe 
and $^{58}$Ni + $^{58}$Ni reactions as a function of $N$ and $Z$ for 30 MeV$/$nucleon beam energy. The solid 
lines are fit to the data as discussed in the text.}
\label{fig:5}       
\end{figure}
\begin{figure}
\resizebox{0.48\textwidth}{!}{
\includegraphics{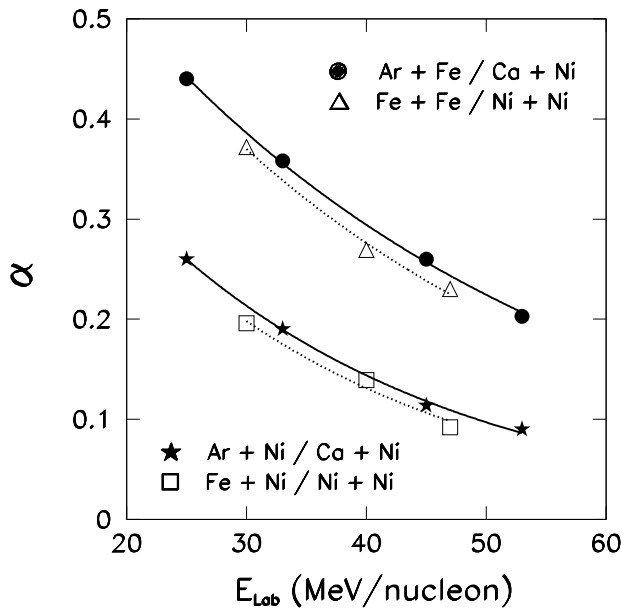}
}
\caption{Experimental isoscaling parameter $\alpha$, as a function of the beam energy. The solid 
         circles are for the  $^{40}$Ar + $^{58}$Fe and $^{40}$Ca + $^{58}$Ni reactions. The open triangles
  are for $^{58}$Fe + $^{58}$Fe and $^{58}$Ni + $^{58}$Ni reactions. The solid stars are for
  $^{40}$Ar + $^{58}$Ni and $^{40}$Ca + $^{58}$Ni reactions. The open squares 
         are for $^{58}$Fe + $^{58}$Ni and $^{58}$Ni + $^{58}$Ni reactions.}
\label{fig:6}       
\end{figure}
\par
In Fig. 4, we show the isotopic yield ratio as a function of neutron number $N$, for Ar + Fe, Ar + Ni and 
Ca + Ni systems at beam energies of 25, 33, 45 and 53 MeV$/$nucleon. The left column shows the ratio for 
the $^{40}$Ar + $^{58}$Fe and $^{40}$Ca + $^{58}$Ni pair of reaction and the right column shows the ratio for 
the $^{40}$Ar + $^{58}$Ni and $^{40}$Ca + $^{58}$Ni pair of reaction. One observes that the ratio for each 
element shows linear behavior in the logarithmic plot and aligns with the neighboring element quite well. 
This feature is observed for all the beam energies and both pairs of reactions studied. One also observes 
that the alignment of the data points varies with beam energies as well as the pairs of reaction. To have a 
quantitative estimate of this variation, the ratio for each element ($Z$) was simultaneously fit using an 
exponential relation (shown by the solid lines) to obtain the slope parameter $\alpha$. The values of the 
parameters are shown at the top of each panel in the figure. The value of the slope parameter $\alpha$ is 
larger for the $^{40}$Ar + $^{58}$Fe and $^{40}$Ca + $^{58}$Ni reactions, which has a larger difference in 
the $N/Z$ of the systems in the pair, compared to the $^{40}$Ar + $^{58}$Ni and $^{40}$Ca + $^{58}$Ni 
reactions, which has a smaller difference in the corresponding $N/Z$. The $\alpha$ value furthermore 
decreases with increasing beam energy. A similar feature is also observed in Fe + Fe, Fe + Ni, Ni + Fe and 
Ni + Ni systems. Fig. 5 shows the isotope yield ratios and the isotone yield ratios for the Fe + Fe and 
Ni + Ni reactions for the 30 MeV$/$nucleon beam energy.  A relative comparison of how the isoscaling 
parameter $\alpha$, evolves as a function of beam energy and the isospin of the system is shown in Fig. 6. 
The figure clearly shows that the $\alpha$ value decreases with beam energy from 25 MeV$/$nucleon to 53 
MeV$/$nucleon. In addition, there is also a clear drop in the $\alpha$ values with the decrease of the 
$N/Z$ values of the system. 

\section{Theoretical Model Comparison}
\subsection{Dynamical AMD model}
The Antisymmetrized Molecular Dynamics (AMD) \cite{ONO03,ONO04} is a microscopic model that simulates the 
time evolution of a nuclear collision. The colliding system in this model is represented in terms of a fully 
antisymmetrized product of Gaussian wave packets. During the evolution, the wave packet centroids move 
according to the deterministic equation of motion. The followed state of the simulation branches 
stochastically and successively into a huge number of reaction channels. The interactions are parameterized 
in terms of an effective force acting between nucleons and the nucleon-nucleon collision cross-sections. The 
advantage of using a dynamical model to study the nuclear equation of state is that it allows one to 
understand the functional form of the density dependence of the symmetry energy at a very fundamental level 
({\it {i.e.}}, from the basic nucleon-nucleon interaction). 
\par
Recently \cite{ONO03}, the fragment yields from heavy ion collisions simulated within the Antisymmetrized 
Molecular Dynamics (AMD) calculation were reported to follow a scaling behavior of the type shown in Eq. 1. 
A linear relation between the isoscaling parameter $\alpha$ and the difference in the isospin asymmetry 
($Z/A)^{2}$ of the fragments (as given in Eq. 2), with appreciably different slopes, was predicted for two 
different forms of the density dependence of the symmetry energy; a `` stiff " dependence (obtained from 
Gogny-AS interaction) and a `` soft " dependence (obtained from Gogny interaction).
\par
In this section, we compare the experimentally determined isoscaling parameter with the predictions of
the AMD model calculation. The isospin asymmetry of the fragments for the present systems was
estimated at {\it {t}} = 300 {\it {fm/c}} of the dynamical evolution using the AMD calculation. The values for the
fragment asymmetry $(Z/A)^{2}$, were obtained by interpolating between those calculated for the 
$^{40}$Ca + $^{40}$Ca, $^{48}$Ca + $^{48}$Ca and $^{60}$Ca + $^{60}$Ca systems by Ono {\it {et al.}} \cite{ONO03}. 
These systems are symmetric and nearly similar in charge and mass as studied in the present work.
\begin{figure}
\resizebox{0.48\textwidth}{!}{
\includegraphics{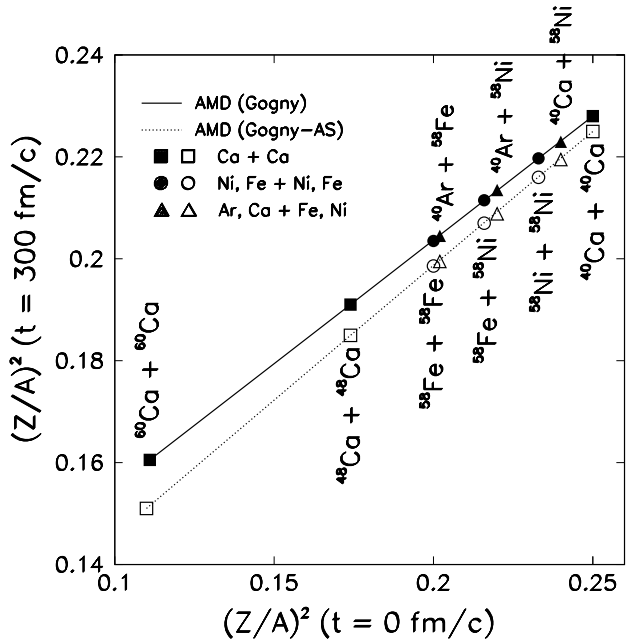}
}
\caption{AMD calculations of the fragment asymmetry (Z/A)$^{2}$, at {\it {t}} = 300 {\it {fm/c}} for the Gogny 
        (solid line and solid squares) and Gogny-AS (dotted line and hollow squares) interactions at 
 35 MeV/nucleon. The calculations are taken from Ref. \cite{ONO03} for the systems shown by the square 
 symbols. The lines are linear fit to the square symbols. The other symbols are the interpolated values 
 for the systems studied in this work.}
\label{fig:7}       
\end{figure}
Fig. 7 shows the AMD calculation of the fragment asymmetry, $(Z/A)^{2}$ at {\it {t}} = 300 {\it {fm/c}}, as a 
function of initial asymmetry at time {\it {t}} = 0 {\it {fm/c}}, for  two different choices of the nucleon-nucleon 
interaction, Gogny and Gogny-AS. The asymmetry values for the $^{40}$Ca + $^{40}$Ca, $^{48}$Ca + $^{48}$Ca 
and $^{60}$Ca + $^{60}$Ca systems of Ref. \cite{ONO03} are shown by solid and hollow square symbols for the 
Gogny and Gogny-AS interaction, respectively. The lines are the linear fits to the calculations. The 
interpolated values for the present systems are shown by the solid circles and triangles, and hollow circles 
and triangles for the Gogny and Gogny-AS interaction, respectively. 
\par
We note that the AMD calculations carried out in Ref. \cite{ONO03} and shown in Fig. 7 correspond to 
the beam energy of 35 MeV/nucleon. The interpolated values of the asymmetries for the present systems 
obtained from Fig. 7 are therefore for the beam energy of 35 MeV/nucleon. In order to compare the 
experimentally determined isoscaling parameter to that of the calculations, we therefore make use of 
the experimental isoscaling parameter for the beam energy of 35 MeV/nucleon using Fig. 6.
\par
Fig. 8 shows  a comparison between the experimentally observed $\alpha$ and those from the
AMD model calculations plotted as a function of the difference in the fragment asymmetry for the beam 
energy of 35 MeV/nucleon. The solid and the dotted lines are the AMD model predictions for the `` soft " 
(Gogny) and the `` stiff " (Gogny-AS) form of the density dependence of the symmetry energy, respectively. 
The solid and the hollow symbols (squares, stars, triangles and circles) are the results of  the present 
study for the two different values of the fragment asymmetry, assuming Gogny and Gogny-AS interactions, 
respectively. Also shown in the figure are the scaling parameters (asterisks, crosses, diamond and inverted 
triangle) taken 
\begin{figure}
\resizebox{0.48\textwidth}{!}{
\includegraphics{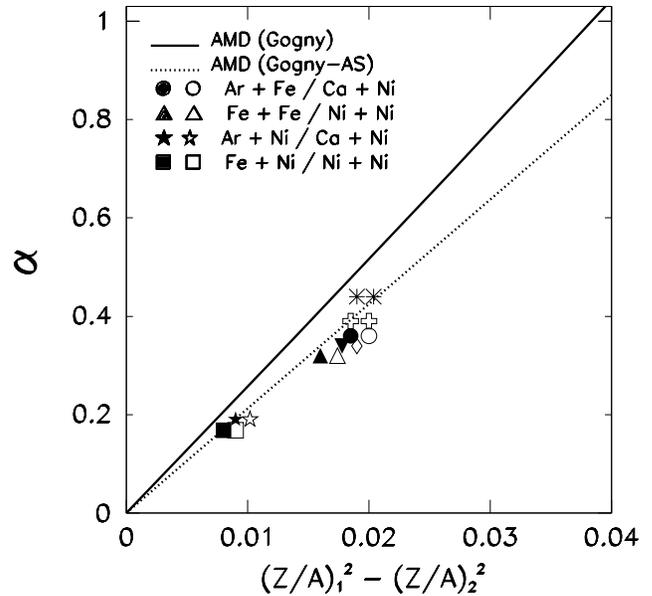}
}
\caption{Isoscaling parameter $\alpha$, as a function of the difference in fragment asymmetry for 35 
         MeV/nucleon. The solid and the dotted lines are the AMD calculations for the Gogny and Gogny-AS 
  interactions, respectively \cite{ONO03}. The solid and the hollow squares, stars, triangles and 
  circles are from the present work as described in the text. The other symbols corresponds to data 
  taken from \cite{GER04} (asterisks) and \cite{BOT02} (crosses, diamonds,
         inverted triangles).}
\label{fig:8}       
\end{figure}
from various other works in the literature \cite{BOT02,GER04}. It is observed that the experimentally 
determined $\alpha$ parameter increases linearly with increasing difference in the fragment asymmetry of the 
two systems as predicted by the AMD calculation. Also, the data points are in closer agreement with those 
predicted by the Gogny-AS interaction (dotted line) than those from the usual Gogny force (solid line). 
\par
In the above comparison between the data and the calculation, the corrections for the 
isoscaling parameter $\alpha$ due to the sequential de-excitation of the fragments are not taken into account. 
The slightly lower values of the isoscaling parameters (symbols) from the present measurements with respect to 
the Gogny-AS values (dotted line) could be due to the small secondary de-excitation effect of the fragments not 
accounted for in this comparison. Recently, it has been reported by Ono {\it {et al.}} \cite{ONO06}, that the 
sequential decay effect in the dynamical calculations can affect the $\alpha$ value by as much as
50 $\%$, and the ability to distinguish between the `` stiff " and the `` soft " form of the density
dependence of the symmetry energy diminishes significantly. The 
calculations by Ono {\it {et al.,}} were carried out for the above studied systems using the AMD model. However, 
dynamical calculation carried out by Tian {\it {et al.}} \cite{TIA06}, using Isospin Quantum Molecular 
Dynamic (IQMD) model shows no significant difference between the primary and the secondary $\alpha$. The 
sequential decay effect from the IQMD calculation was also carried out for the same systems and beam energy 
as studied by Ono {\it {et al.}} \cite{ONO06} using the AMD model. The contrasting results between the two 
dynamical calculations for the same systems and energy currently present significant amount of uncertainty in 
reliably estimating the effect of sequential decay using dynamical models. One reason for this could be the 
large discrepancy that exists in the determination of the primary fragment excitation energy from the current 
dynamical models. It has been shown using another dynamical model (stochastic mean field calculation) (see 
Liu {\it {et al.}} \cite{LIU04}), that it requires a significantly lower value of the primary fragment 
excitation energy (by as much as 50$\%$), to be able to reproduce the experimentally observed fragment isotope 
distribution. 
\par
In the above comparison between the data and the calculation, we have therefore assumed the effect of the 
sequential decay to be negligible. A correction of about 10 - 15 $\%$, as determined and well established 
from various statistical model studies \cite{TSANG01}, results in a slight increase in the $\alpha$ values 
bringing them even closer to the dotted line. The observed agreement of the experimental data with the 
Gogny-AS type of interaction therefore appears to suggest a `` stiff " form of the density dependence of 
the symmetry energy. 
\par
Figure 9 shows various forms of the density dependence of the symmetry energy in isospin asymmetric nuclear 
matter used by Chen {\it {et al.}} \cite{CHE05}, and those used in the present dynamical model analysis. The 
dot-dashed, dotted and the dashed curve corresponds to the momentum dependent Gogny interactions used by 
Chen {\it {et al.,}} to explain the NSCL-MSU isospin diffusion data. Assuming that the density dependence of 
the symmetry energy can be parametrized as,
 
\begin{equation}
         C_{sym}(\rho) = C_{sym}^{o} \bigg (\frac{\rho}{\rho_o} \bigg )^{\gamma} (MeV)
\end{equation} 

where $C_{sym}^{o}$, is the value of the symmetry energy at saturation density and $\gamma$ is the parameter 
that characterizes the stiffness of the symmetry energy, the above dependences used by Chen {\it {et al.}} 
can be written as, E$_{sym}$ $\approx$ 31.6 ($\rho/\rho_{\circ})^{\gamma}$, where, $\gamma$ = 1.6, 1.05 and 
0.69, respectively. The solid curve and the solid point in Fig. 9 correspond to those from the Gogny and 
Gogny-AS interactions used to study the isoscaling data in the present work. 
\par
By parameterizing the density dependence of the symmetry energy that explains the present isoscaling data, 
one obtains, C$_{sym}(\rho)$ $\approx$ 31.6 ($\rho/\rho_{\circ})^{\gamma}$, where $\gamma$ = 0.69, from the
dynamical model analysis. 
\begin{figure}
\resizebox{0.48\textwidth}{!}{
\includegraphics{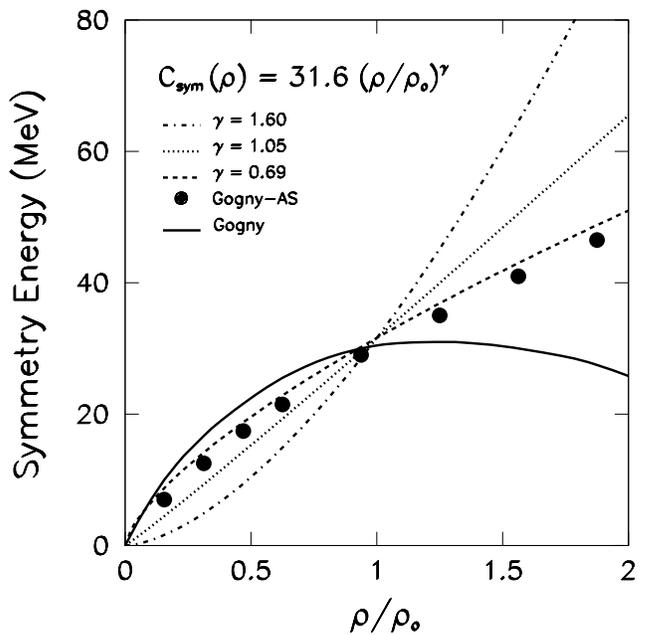}
}
\caption{Different forms of the density dependence of the nuclear symmetry energy 
         used in the dynamical analysis of the present measurements on isoscaling data and the isospin 
  diffusion measurements of NSCL-MSU \cite{CHE05}. The curves are as described in the text.}
\label{fig:9}       
\end{figure}

\subsection{Statistical Multifragmentation Model}

The Statistical Multifragmentation Model (SMM) \cite{BON95,BOT95} is based on the assumption of statistical 
equilibrium at a low density freeze-out stage. All breakup channels composed of nucleons and excited 
fragments are taken into account and considered as partitions.  During each partition the conservation of 
mass, charge, energy, momentum and angular momentum is taken into account, and the partitions are sampled 
uniformly in the phase space according to their statistical weights using Monte Carlo sampling. The Coulomb 
interaction between the fragments is treated in the Wigner-Seitz approximation. Light fragments with mass 
number $A$ $\leq$ 4 are considered as elementary particles with only translational degrees of freedom 
(``nuclear gas"). Fragments with $A$ $>$ 4 are treated as heated nuclear liquid drops, and their individual 
free energies $F_{A,Z}$ are parametrized as a sum of the volume, surface, Coulomb and symmetry energy.
\par
For the present study we make use of the SMM version adopted by Botvina {\it {et al.}} \cite{BOT02}.
In this version, the secondary de-excitation of large fragments with $A$ $>$ 16 is described by Weisskopf-type
evaporation and Bohr-Wheeler-type fission models \cite{BON95,BOT87}. The decay of smaller fragments is 
treated with the Fermi-breakup model. All ground and nucleon-stable excited states of light fragments are
taken into account and the population probabilities of these states are calculated according to the
available phase space \cite{BOT87}. The sequential decay effect on the isoscaling parameter in this
version of SMM has been established to be small and in good agreement with other versions of the statistical 
models. 
\par
Unlike dynamical calculations, the form of the density dependence of the symmetry energy is not known
a priori, but has to be deduced from the systematic correlations between the isoscaling
parameter, temperature, symmetry energy and the density of the multifragmenting system. To build this
correlation, we make use of the fragment yield distributions measured in $^{58}$Ni, $^{58}$Fe + $^{58}$Ni, 
$^{58}$Fe reactions at 30, 40 and 47 MeV/nucleon to study the isoscaling parameter $\alpha$, as a function 
of the excitation energy of the fragmenting source. The parameter $\alpha$ was obtained from
the ratio's of the isotopic yields for two different pairs of reactions, $^{58}$Fe + $^{58}$Ni and $^{58}$Ni +
$^{58}$Ni, and  $^{58}$Fe + $^{58}$Fe and $^{58}$Ni + $^{58}$Ni as discussed in section 4.2. 
The excitation energy of the source for each beam energy was determined by simulating the initial stage of 
the collision dynamics using the Boltzmann-Nordheim-Vlasov (BNV) model calculation \cite{BAR02}.  The results 
were obtained at a time around 40 - 50 fm/c after the projectile had fused with the target nuclei and the 
quadrupole moment of the nucleon coordinates (used for identification of the deformation of the system) 
approached zero. These excitation energies were also compared with those obtained from the systematic 
calorimetric measurements (see
\begin{figure}
\resizebox{0.48\textwidth}{0.72\textheight}{
\includegraphics{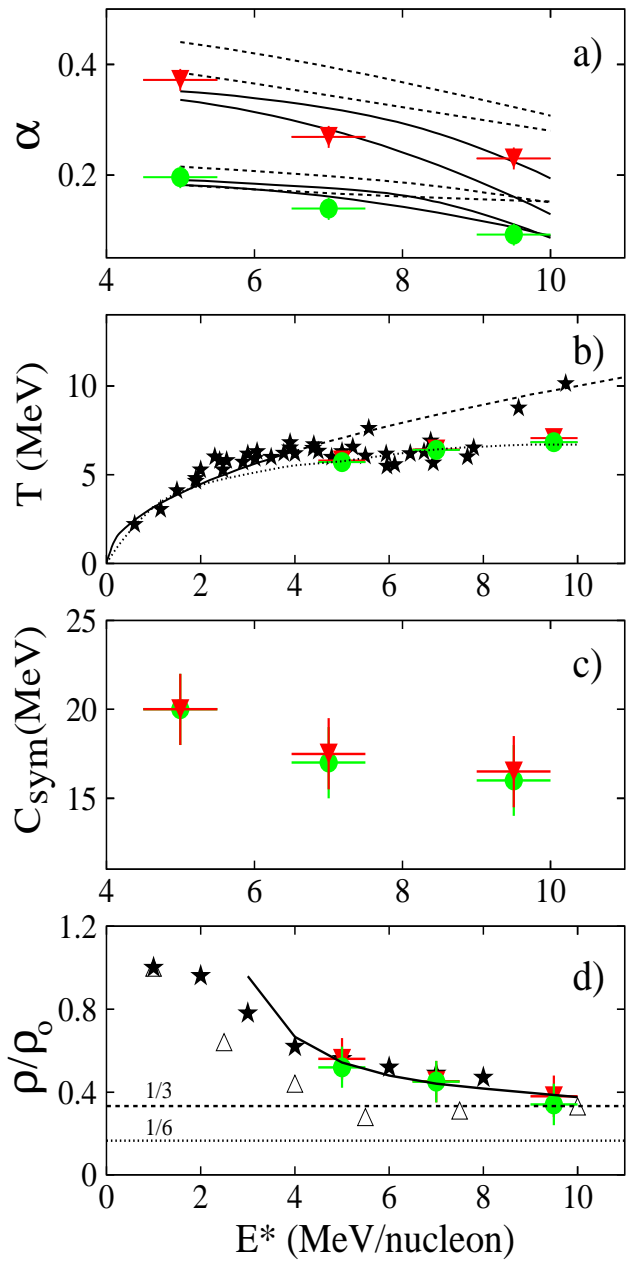}
}
\caption{(Color online) Isoscaling parameter $\alpha$, temperature, symmetry energy and density 
as a function of excitation energy for the Fe + Fe and Ni + Ni (inverted triangles), and Fe + Ni and Ni + Ni
(solid circles) reactions at 30, 40 and 47 MeV/nucleon. a) Experimental isoscaling parameter as a function of 
excitation energy. The solid and the dashed curves are the SMM calculations as discussed in the text. b)
Temperature as a function of excitation energy. The solid stars correspond to the measured values and are
taken from Ref. \cite{NAT02}. The solid and the dashed curve corresponds to the Fermi-gas relation. The
dotted curve corresponds to the one obtained from Eq. 4. c) Symmetry energy as a function of excitation 
energy. d) Density as a function of excitation energy. The solid stars correspond to those from 
Ref. \cite{NATO02}. The open triangles are those from Ref. \cite{VIO04}. The solid curve is from 
Ref. \cite{BON85}.}
\label{fig:10}       
\end{figure}
Ref. \cite{NAT02}) for systems with mass ($A$ $\sim$ 100), and similar to those studied in the present work, 
and are in good agreement. Fig. 10 (a) shows the experimental isoscaling parameter $\alpha$, as a function 
of the excitation energy for Fe + Fe and Ni + Ni, and Fe + Ni and Ni + Ni pairs of reactions. A systematic 
decrease in the absolute values of the isoscaling parameter with increasing excitation energy is observed 
for both pairs. The $\alpha$ parameters for the $^{58}$Fe + $^{58}$Fe and $^{58}$Ni + $^{58}$Ni are about 
twice as large compared to those for the $^{58}$Fe + $^{58}$Ni and $^{58}$Ni + $^{58}$Ni pair of reactions. 
\par
The experimental isoscaling parameter was compared with the predictions of the Statistical Multifragmentation 
Model (SMM) \cite{BON95,BOT01} calculations to study their dependence on the excitation energy and the
isospin content.  The initial parameters such as, the mass, charge and excitation energy of the fragmenting 
source for the calculation was obtained from the BNV calculations as discussed above. The possible 
uncertainties in the source parameters due to the loss of nucleons during pre-equilibrium emission was 
accounted for by carrying out the calculations for smaller source sizes. The break-up density in the 
calculation was taken to be multiplicity-dependent and was varied from approximately 1/2 to 1/3 the 
saturation density. This was achieved by varying the free volume with the excitation energy as shown in 
Ref. \cite{BON95}. The form of the dependence was adopted from the work of Bondorf {\it {et al.}} \cite{BON85,BON98}, 
(and shown by the solid curve in Fig. 10 (d)). It is known that the multiplicity-dependent break-up density, 
which corresponds to a fixed interfragment spacing and constant pressure at break-up, leads to a pronounced 
plateau in the caloric curve \cite{BON85,BON98}. A constant break-up density would lead to a steeper 
temperature versus excitation energy dependence. 
\par
The symmetry energy in the calculation was varied until a reasonable agreement between the calculated and the 
measured $\alpha$ was obtained. Fig. 10 (a) shows the comparison between the SMM calculated and the 
measured $\alpha$ for both pairs of systems. The dashed curves correspond to the calculation for the primary 
fragments and the solid curves to the secondary fragments. The width in the curve is the measure of the 
uncertainty in the inputs to the SMM calculation. It is observed that, within the given uncertainties,  the 
decrease in the $\alpha$ values with increasing excitation energy and decreasing isospin difference 
$\Delta(Z/A)^2$, of the systems is well reproduced by the SMM calculation. One also notes that the effect of 
sequential decay on the isoscaling parameter is small as observed in several other studies \cite{TSA01,TAN01} 
using statistical models. 
\par
We show in Fig. 10 (b), the temperature as a function of excitation energy ({\it {caloric curve}}) obtained 
from the above SMM calculation that uses the excitation energy dependence of the break-up density to explain 
the observed isoscaling parameters. These are shown by the solid and inverted triangle symbols. Also shown in 
the figure are the experimentally measured caloric curve data compiled by Natowitz {\it {et al.}} \cite{NAT02}, 
from various measurements for this mass range. The data from these measurements are shown collectively by 
solid star symbols and no distinction is made among them. The Fermi-gas model predictions with inverse level 
density parameter $K_{o}$ = 10 (solid and dashed curve), is also shown. It is evident from the figure that 
the temperatures obtained from the SMM calculations are in good agreement with the overall trend of the 
caloric curve. Somewhat lower value for the temperature is observed when the break-up density of the system 
is kept constant at 1/3 the normal nuclear density. By allowing the break-up density to evolve with the 
excitation energy, a near plateau that agrees with the experimentally measured caloric curves is obtained. 
This assures that the input parameters used in the SMM calculation for comparing with the data are reasonable.
\par
The symmetry energies obtained from the statistical model comparison of the experimental isoscaling
parameter $\alpha$, are as shown in Fig. 10 (c). A steady decrease in the symmetry energy with increasing 
excitation energy is observed for both pairs of systems. Such a decrease has also been observed in several 
other studies \cite{IGL06,SHE05,FEV05,HEN05}. We have also estimated the effect of the symmetry energy 
evolving during the sequential de-excitation of the primary fragments \cite{IGL06,BUY05}. These are reflected 
in the large error bars shown in Fig. 10 (c).
\par
\begin{figure}
\resizebox{0.48\textwidth}{!}{
\includegraphics{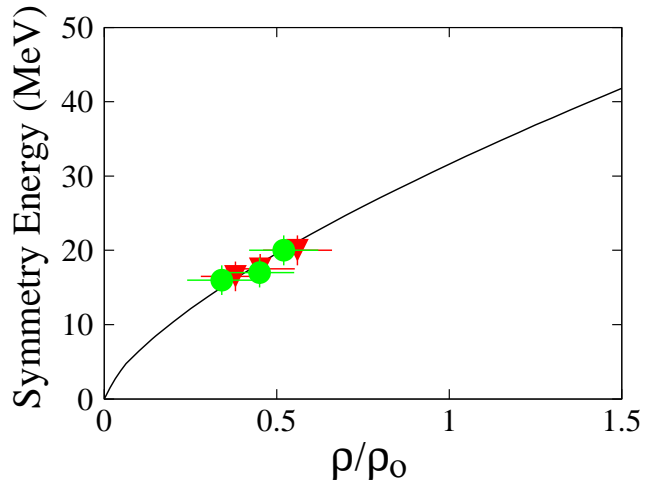}
}
\caption{(Color online) Symmetry energy as a function of density for the Fe + Fe and Ni + Ni pair of
    reaction (inverted triangles), and Fe + Ni and Ni + Ni pair of reactions (solid circles) for the 30, 40 
    and 47 MeV/nucleon.  The solid curve is the dependence obtained form the dynamical model analysis as 
    explained in the text.}
\label{fig:11}       
\end{figure}

The phase diagram of the multifragmenting system is two dimensional and hence the excitation energy
dependence of the temperature (the caloric curve) must take into account the density dependence too.
Often this dependence is neglected while studying the caloric curve. In the following, we attempt to extract 
the density of the fragmenting system as a function of excitation energy. It has been shown by 
Sobotka {\it {et al.}} \cite{SOB04}, that the plateau in the caloric curve could be a consequence of 
the thermal expansion of the system at higher excitation energy and decreasing density. By assuming that the 
decrease in the breakup density, as taken in the present statistical multifragmentation calculation, can be 
approximated by the expanding Fermi gas model, and furthermore the temperature in Eq. 2 and the temperature 
in the Fermi-gas relation are related, one can extract the density as a function of excitation energy using 
the relation

\begin{equation}
      T = \sqrt {K_{o} (\rho / \rho_o)^{2/3} E^*}
\end{equation}

In the above expression, the momentum and the frequency dependent factors in the effective mass ratio are 
taken to be one as expected at high excitation energies and low densities studied in this work
\cite{HAS86,SHL90,SHL91}.
\par
The resulting densities for the two pairs of systems are shown in Fig. 10 (d) by the solid circles and 
inverted triangles. For comparison, we also show the break-up densities obtained from the analysis of the 
apparent level density parameters required to fit the measured caloric curve by 
Natowitz {\it {et al.}} \cite{NATO02}, and those obtained by Viola {\it {et al.}} \cite{VIO04} from the 
Coulomb barrier systematics that are required to fit the measured intermediate mass fragment kinetic energy 
spectra. One observes that the present results obtained by requiring to fit the measured isoscaling 
parameters and the caloric curve are in good agreement with those obtained by Natowitz {\it {et al}}. The 
\begin{table*}
\caption{Parameterized form of the density dependence of the symmetry energy obtained
from various independent studies.}
\label{tab:1}       
\begin{tabular}{lllllllllllllllll}
\hline\noalign{\smallskip}
Reference &  & & & &  & & & Parametrization & & & & &  & & & Studies  \\
\noalign{\smallskip}\hline\noalign{\smallskip}
Fuchs {\it {et al.}}       \cite{FUCH06} & & & &  & & & &  32.9($\rho$/$\rho_{o}$)$^{0.59}$          & &  & & & & & & Relativistic Dirac-Brueckner calculation  \\
Heiselberg {\it {et al.}}  \cite{HEI00}  & & & &  & & & &  32.0($\rho$/$\rho_{o}$)$^{0.60}$          & &  & & & & & & Variational calculation             \\
Danielewicz {\it {et al.}} \cite{DAN04}  & & & &  & & & &  31(33)($\rho$/$\rho_{o}$)$^{0.55(0.79)}$  & &  & & & & & & BE, skin, isospin analog states     \\
Tsang {\it {et al.}}       \cite{TSA04}  & & & &  & & & &  12.125($\rho$/$\rho_{o}$)$^{2}$           & &  & & & & & & Isospin diffusion                   \\
Chen {\it {et al.}}        \cite{CHE05}  & & & &  & & & &  31.6($\rho$/$\rho_{o}$)$^{1.05}$          & &  & & & & & & Isospin diffusion                   \\  
Li {\it {et al.}}          \cite{BAL05}  & & & &  & & & &  31.6($\rho$/$\rho_{o}$)$^{0.69}$          & &  & & & & & & Isospin diffusion                   \\  
Piekarewicz {\it {et al.}} \cite{PIE05,PIE06}   & & & &  & & & &  32.7($\rho$/$\rho_{o}$)$^{0.64}$   & &  & & & & & & Giant resonances                    \\ 
Shetty {\it {et al.}}      \cite{SHE04,SHET05,SHET06}   & & & &  & & & &  31.6($\rho$/$\rho_{o}$)$^{0.69}$  & &  & & & & & & Isotopic distribution        \\     
Famiano {\it {et al.}}     \cite{FAM06}      & & & &  & & & &  32.0($\rho$/$\rho_{o}$)$^{0.55}$         & &  & & & & & & neutron-proton emission ratio    \\  
Tsang {\it {et al.}}       \cite{TSANG01}    & & & &  & & & &  23.4($\rho$/$\rho_{o}$)$^{0.6}$          & &  & & & & & & Isotopic distribution            \\     
\noalign{\smallskip}\hline
\end{tabular}
\end{table*}
figure also shows the fixed freeze-out density of 1/3 (dashed line) and 1/6 (dotted line) of the saturation 
density assumed in various statistical model comparisons. The caloric curve obtained using the above 
densities and excitation energies (shown by solid stars, circles and the triangles) with $K_{o}$ = 10 in 
Eq. 4, is shown by the dotted curve in Fig. 10 (b). The small discrepancy between the dotted curve and the 
data (solid stars) below 4 MeV/nucleon is due to the approximate nature of Eq. 4 being used.   
\par
It is therefore evident from figure 10 (a), (b), (c) and (d) that the decrease in the experimental isoscaling 
parameter $\alpha$, symmetry energy, break-up density, and the flattening of the temperature with increasing 
excitation energy are all correlated. One can thus conclude that the expansion of the system during the 
multifragmentation process leads to a decrease in the isoscaling parameter, decrease in the symmetry energy 
and density, and the flattening of the caloric curve. 
\par
From the above correlation between the symmetry energy as a function of excitation energy, and the density 
as a function of excitation energy, we obtain the symmetry energy as a function of density. This is shown
by the inverted triangles and solid circles in Fig. 11 for the Fe + Fe and Ni + Ni, and the Fe + Ni and 
Ni + Ni pair of reactions. The temperature in the present work remains nearly constant for the range of 
excitation energies studied, the  observed decrease in the symmetry energy with increasing excitation energy 
is therefore a consequence of decreasing density. This is also supported by microscopic calculations which 
shows an extremely slow evolution of the symmetry energy with temperature \cite{BAL01,BAL06}. The evolution 
is practically negligible for the temperature range studied in this work.  The solid curve in Fig. 11 
corresponds to the dependence C$_{sym} (\rho)$ $=$ 31.6 ($\rho/\rho_{\circ})^{0.69}$ MeV, obtained from the 
dynamical Antisymmetrized Molecular Dynamic (AMD) calculation, as discussed in the previous section. It is
thus observed that the dynamical and statistical models lead to similar density dependence of the symmetry
energy.

\section{Comparison with other independent studies}

In the following, we compare the form of the density dependence of the symmetry energy obtained from the 
present experimentally measured isoscaling parameter using the statistical and the dynamical multifragmentation 
models with several other recent independent studies. Fig. 12 shows this comparison. The green solid curve 
corresponds to the one obtained from Gogny-AS interaction in dynamical AMD model that explains the present 
results \cite{SHE04,SHET05}, assuming the sequential decay effect to be small. The inverted triangle and the 
circle symbols also correspond to the present measurements obtained by comparing with the statistical 
multifragmentation model \cite{SHET06}. The red dashed curve corresponds to the one obtained recently from 
an accurately calibrated relativistic mean field interaction, used for describing the Giant Monopole 
Resonance (GMR) in $^{90}$Zr and $^{208}$Pb, and the IVGDR in $^{208}$Pb by 
Piekarewicz {\it {et al}}. \cite{TOD05,PIE05,PIE06}. The pink dot-dashed curve correspond to the one used 
to explain the isospin diffusion results of NSCL-MSU using the isospin dependent Boltzmann-Uehling-Uhlenbeck 
(IBUU) model by Tsang {\it {et al.}} \cite{TSA04}. The blue dot-dashed curve also corresponds to the one 
used for explaining the isospin diffusion data of NSCL-MSU by Chen {\it {et al.}} \cite{CHE05}, but with the 
momentum dependence of the interaction included in the IBUU calculation. This dependence has been further 
modified to include the isospin dependence of the in-medium nucleon-nucleon cross-section by 
Li {\it {et al.}} \cite{BAL05}, and is in good agreement with the present study. The shaded region 
in the figure corresponds to those obtained by constraining the binding energy, neutron skin thickness and 
isospin analogue state in finite nuclei using the mass formula of Danielewicz \cite{DAN04}. The yellow solid 
curve correspond to the parameterization adopted by Heiselberg {\it {et al.}} \cite{HEI00} in their studies 
\begin{figure*}
\resizebox{0.65\textwidth}{!}{
\includegraphics{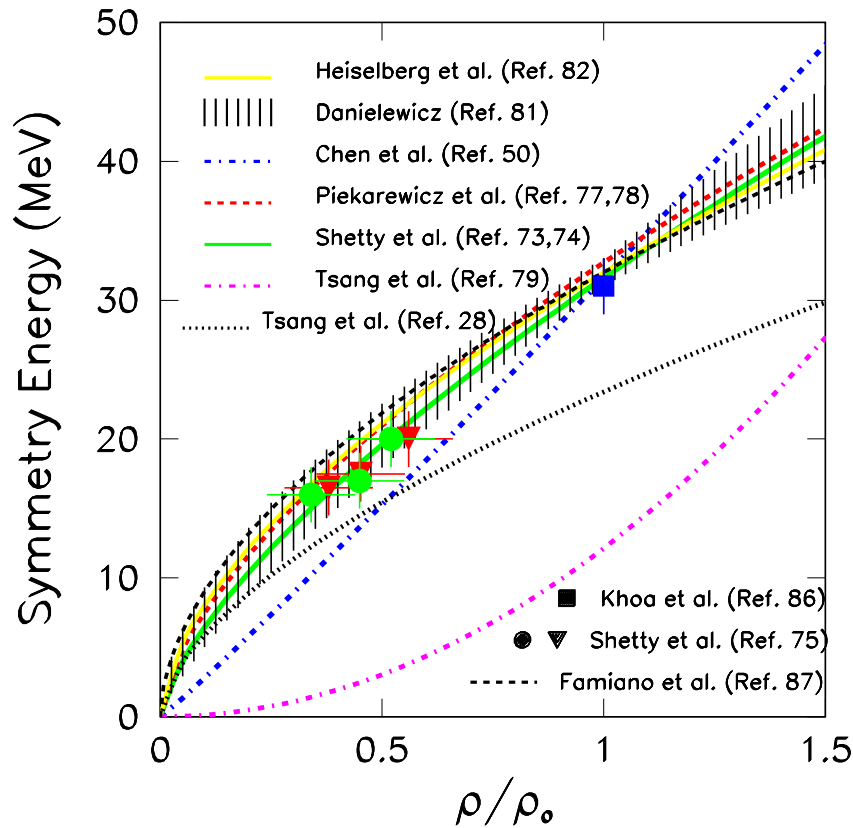}
}
\caption{(Color online) Comparison between the results on the density dependence of the symmetry energy
obtained from various different studies. The various curves and the symbols are described in the text.}
\label{fig:12}       
\end{figure*}
on neutron stars. By fitting earlier predictions of the variational calculations by 
Akmal {\it {et al.}} \cite{AKM97,AKM98}, where the many-body and special relativistic corrections are 
progressively incorporated, Heiselberg and Hjorth-Jensen obtained a value of C$_{sym}^{o}$ = 32 MeV and 
$\gamma$ = 0.6, similar to those obtained from the present measurements. A similar result is also obtained
from the relativistic Dirac-Brueckner calculation, with C$_{sym}^{o}$ = 32.9 MeV and 
$\gamma$ = 0.59 \cite{FUCH06}. The Dirac-Brueckner is an ``ab-initio" calculation based on nucleon-nucleon 
interaction with Bonn A type potential instead of the AV18 potential used in the variational calculation of 
Ref. \cite{AKM98}. The density dependence of the symmetry energy has also been studied in the framework of 
expanding emitting source  (EES) model by Tsang {\it {et al.}} \cite{TSANG01}, where a power law dependence 
of the form C$_{sym}(\rho)$ = 23.4($\rho/\rho_{o})^{\gamma}$, with $\gamma$ = 0.6 was obtained. This 
dependence (shown by the black dotted curve) is significantly softer than other dependences shown in 
the figure. The solid square point in the figure correspond to the value of symmetry energy obtained by 
fitting the experimental differential cross-section data in a charge exchange reaction using the isospin 
dependent CDM3Y6 interaction of the optical potential by Khoa {\it {et al.}} \cite{KHO05}. 
\par
An alternate observable, the double neutron/proton ratio of nucleons taken from two reaction
systems using four isotopes of the same element, has recently been proposed as a probe to study the
density dependence of the symmetry energy \cite{FAM06}. This observable is expected to be more robust than
the isoscaling observable. It was shown recently \cite{FAM06} that the experimentally determined 
double-ratio for the $^{124}$Sn + $^{124}$Sn reaction to that for the  $^{112}$Sn + $^{112}$Sn
reaction, results in a dependence with $\gamma$ = 0.5 (shown by black dashed curve in Fig. 12), when compared 
to the predictions of the IBUU transport model calculations. This observation is in
close agreement with other studies discussed above. However, this dependence has been obtained by using
the momentum independent calculation of Ref. \cite{BAL97}. A more recent calculation \cite{BALI06} using a
BUU transport model that includes momentum dependent interaction show significantly lower values for the 
double neutron/proton ratio of free nucleons compared to the one reported by Famiano {\it {et al.}}
\par
The parameterized forms of the density dependence of the symmetry energy obtained from all the above
mentioned studies are as given in Table I. The  close agreement between various independent studies show 
that a constraint on the density dependence of the symmetry energy, given 
as C$_{sym}(\rho)$ = C$_{sym}^{o}$($\rho/\rho_{o})^{\gamma}$, where C$_{sym}^{o}$ $\approx$ 31 - 33 MeV 
and $\gamma$ $\approx$ 0.55 - 0.69 can be obtained.

\section{Discussion}

We make the following observations from the above comparison between the statistical and the dynamical model
analysis :

1) {\it {Assuming a negligibly small sequential decay effect, the form of the density dependence of the
symmetry energy obtained from the dynamical model analysis is in good agreement with the one obtained from 
the statistical model analysis}}:  As mentioned earlier, the sequential decay effect among various dynamical 
model calculations is still a subject of debate. The statistical models however consistently show small sequential decay 
effect. If the sequential decay in both the dynamical and the statistical model is determined by the excitation 
energy, charge ($Z$) and mass ($A$) of the fragments, and not by the process that leads to these fragments, the
de-excitation of the fragments 
must lead to a same amount of change in the isoscaling parameter (either a large change or no 
change at all). It is therefore unrealistic to assume that the sequential decay effect is different in the dynamical 
and the statistical model calculations.   One comparison by Hudan {\it {et al.}} \cite{HUD03}, 
show good agreement between the experimentally determined primary fragment excitation energy and those 
calculated using the statistical multifragmentation model (see table II of Ref. \cite{HUD03}). Furthermore, if 
dynamical and statistical models are merely two different ways of interpreting the same multifragmentation
process ({\it {i.e.,}} one simulating the entire process from the formation to the breakup stage, and the other 
simulating only the later breakup stage),  the isoscaling parameter from both interpretation must lead to
consistent results.  It is well known, and as discussed in section II, that both interpretations predict isoscaling 
in multifragmentation. As discussed in section V A, the apparent disagreement between the
sequential decay effect in statistical and dynamical models, could be due to the large discrepancy that exists
in the determination of the primary fragment excitation energy from current dynamical models.
\par
It has been argued \cite{COL06} that the effect of sequential decay on the isoscaling parameter $\alpha$,
in statistical multifragmentation model depends not only on the excitation energy but also on the value of 
the symmetry energy. The fragments in their primary stage  are usually hot and the properties of hot 
nuclei ({\it {i.e.,}} their binding energy and mass) differ from those of cold nuclei. If hot fragments in 
the freeze-out configuration have smaller symmetry energy, their mass at the beginning of the sequential 
de-excitation will be different and this effect should be taken into account. At smaller values of the 
symmetry energy the sequential decay effect can be large. In order to study the effect of symmetry energy 
evolution on isoscaling parameter during sequential decay, we have adopted in this work a phenomenological approach of 
Buyukcizmeci {\it {et al.}} \cite{BUY05}. Fig. 13 and 14 shows the primary and the secondary isoscaling parameter as 
a function of symmetry energy calculated from the statistical multifragmentation model (SMM) for the Ar + Ni 
and Ca + Ni, and Ar + Fe and Ca + Ni pair of reactions. The various panels from top to bottom correspond to
different system excitation energies. Fig. 13 shows the result of the calculations where the symmetry energy is kept 
fixed, and Fig. 14 shows the result for the calculations where the symmetry 
energy is varied during the de-excitation process. The dashed lines in the figure correspond to the primary 
fragments (Eq. 2) and the solid lines to the secondary fragments. It is observed that there is no significant 
change in the primary and the secondary alpha.
\begin{figure}
\resizebox{0.48\textwidth}{!}{
\includegraphics{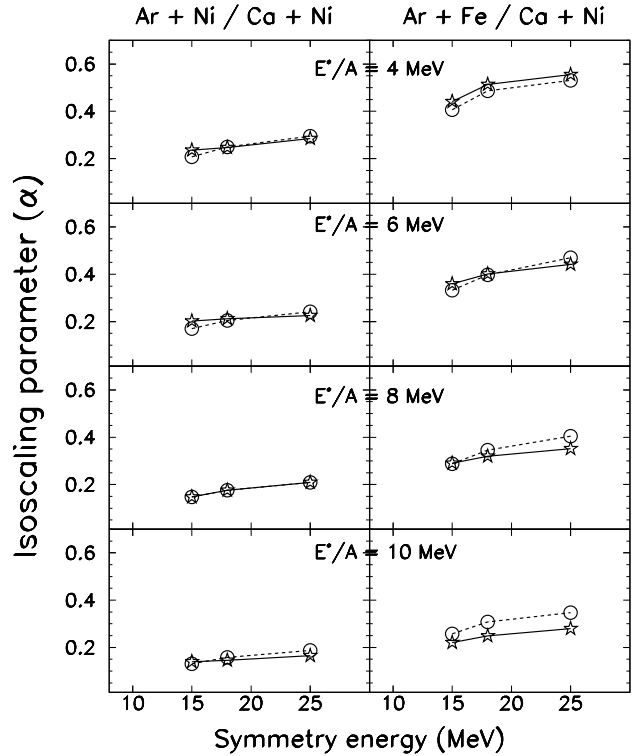}
}
\caption{SMM calculated isoscaling parameter $\alpha$ as a function of symmetry energy for various 
       excitation energies. The open circles joined by dotted lines correspond to the primary
       fragments and the open stars joined by solid lines to the secondary
      fragments. The left column shows the calculation for $^{40}$Ar + $^{58}$Ni and $^{40}$Ca +
      $^{58}$Ni pair, and the right column for the $^{40}$Ar + $^{58}$Fe and $^{40}$Ca + $^{58}$Ni pair.}
\label{fig:13}       
\end{figure}
\begin{figure}
\resizebox{0.48\textwidth}{!}{
\includegraphics{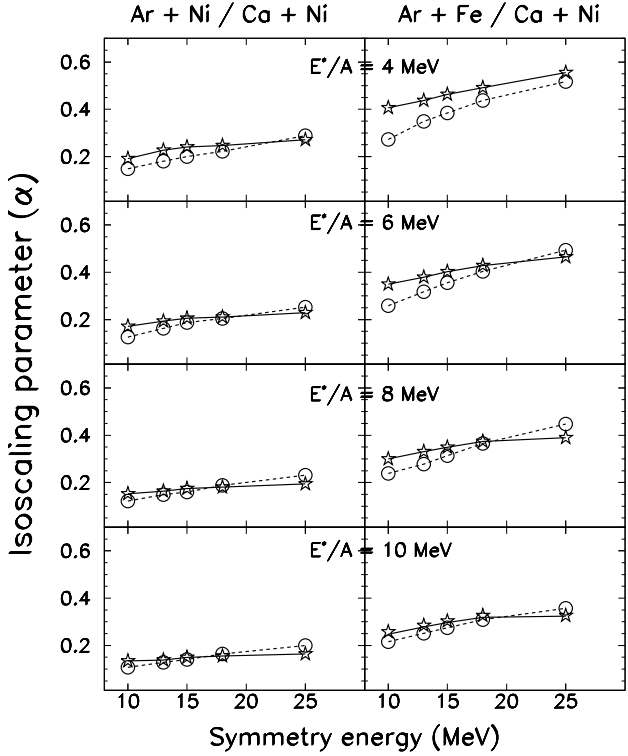}
}
\caption{Same as in fig. 13, but with the modified secondary de-excitation with evolving symmetry energy.}
\label{fig:14}       
\end{figure}
\par
2) {\it {The result of the statistical model analysis is in good agreement with other independent studies}} : 
A comparison between the density dependence of the symmetry energy obtained from the statistical model 
analysis (for which the sequential decay effect is known to be small) and other independent studies shows 
excellent agreement. 
\par
3) {\it {The isoscaling parameter probes the property of infinite nuclear matter}} : The symmetry energy 
obtained from dynamical model analysis (shown by the solid curve in Fig. 11) relates to the volume part of 
the symmetry energy as in infinite nuclear matter, whereas, the symmetry energy obtained from the statistical 
model analysis (solid circles and inverted triangles in Fig. 11) relates to the fragment that is finite and 
has surface contribution. The similarity between the two can probably be understood in terms of the weakening 
of the surface symmetry free energy when the fragments are being formed. During the density fluctuation in 
uniform low density matter, the fragments are not completely isolated and continue to interact with each 
other, resulting in a decrease in the surface contribution as predicted by Ono {\it {et al.}} \cite{ONO04}. 
It must be mentioned that by fitting the nuclear masses with mass formula, a volume contribution to the 
symmetry energy of about 31 - 33 MeV and surface contribution of about 11 - 13 MeV was obtained by 
Danielewicz \cite{DAN03} for nuclei at normal density. Using the constraint obtained for the volume part of 
the symmetry energy from the present study, and following the expression for the symmetry energy of 
finite nuclei by Danielewicz, we write the general expression for the density dependence of the symmetry
energy as,
\begin{equation}
  S_A(\rho) = \frac{\alpha(\rho/\rho_{\circ})^{\gamma}}{1 + [\alpha(\rho/\rho_{\circ})^{\gamma}/\beta A^{1/3}]}
\end{equation}

where, $\alpha$ $\equiv$ $C_{sym}^{o}$ = 31 - 33 MeV, $\gamma$ = 0.55 - 0.69 and $\alpha/\beta$ = 2.6 - 3.0. 
The quantities $\alpha$ and $\beta$ are the volume and the surface symmetry energy at
normal nuclear density. The above equation reduces to Eq. 3 for infinite nuclear matter in the limit of A
$\rightarrow$ $\infty$, and to the symmetry energy of finite nuclei for $\rho = \rho_{\circ}$.
The ratio of the volume symmetry energy to the surface symmetry energy ($\alpha/\beta$) is closely related to 
the neutron skin thickness. Depending upon how the nuclear surface and the Coulomb contribution is treated, 
two different correlations between the volume and the surface symmetry energy have been predicted \cite{STE05} 
from fits to nuclear masses. Experimental masses and neutron skin thickness measurements for nuclei with 
$N/Z$ $>$ 1 should provide tighter constraint on the surface-volume correlation.
\par
4) {\it {The density dependence of the symmetry energy obtained using the statistical model approach is 
consistent with other experimentally determined observables}} : In the past, attempts have been made to 
study the density dependence of the symmetry energy by looking at specific observables and comparing them 
with the predictions of the dynamical models. Such an approach attempts to explain the observable of 
interest  without trying to simultaneously explain other properties, such as, the temperature, density and 
excitation of the fragmenting system. This has lead to a variety of different dependences without any clue 
to what density is being probed. It might not be straightforward to distinguish different realistic EOS 
interactions using dynamical models, due to the large uncertainties that currently exist in the sequential 
decay effects for these models. But  the idea of extracting information on the symmetry energy  from the 
point of view of the basic nucleon-nucleon interaction is a very powerful approach. On the other hand, the
determination of the density dependence of the symmetry energy from statistical model analysis by 
simultaneously explaining the isoscaling parameter, caloric curve and the density as a function of excitation 
energy is a reverse approach. This approach attempts to explain  the experimental observables without any 
prior knowledge of the governing interaction  and arrives at a dependence which can then be compared with 
those predicted from the basic interactions.
\par
5) {\it {Symmetry energy determined from the multifragmentation study is lower than that of normal
nuclei}} : Theoretical many-body calculations \cite{DIE03,ZUO99,BRA85,PEA00} and those from the empirical 
liquid-drop mass formula \cite{MYE66,POM03} predict symmetry energy near normal nuclear density and 
temperature to be around 30 - 32 MeV. Assuming a negligible evolution of the symmetry energy as a function of 
temperature, as shown in Ref. \cite{BAL01,BAL06}, the present statistical model analysis yields symmetry 
energy of the order of 18 - 20 MeV at half the normal nuclear density.  
\par
6) {\it {The above constraint on the density dependence of the symmetry energy has important implications for 
astrophysical and nuclear physics studies :}} 
\par
a) {\it {Neutron skin thickness}}:
It has been shown \cite{HOR06} that an empirical fit to a large number of mean-field calculations yield 
neutron skin thickness for $^{208}$Pb nucleus, $R_{n}$ - $R_{p}$ $\simeq$ (0.22$\gamma$ + 0.06) {\it {fm}}, 
where $\gamma$ is the exponent that determines the stiffness of the density dependence of the symmetry 
energy. From the above comparison, assuming only those density dependences of the symmetry energy which have 
symmetry energy at normal nuclear density 31 - 33 MeV,  one obtains a neutron skin thickness of 
0.18 - 0.21 {\it {fm}}. An accurate determination of the neutron skin thickness from the parity-violating 
electron scattering measurement \cite{HORO01} will provide a unique observational constraint on the thickness 
of the neutron skin of a heavy nucleus. The above constraint also leads to symmetric matter nuclear 
compressibility of $K$ $\sim$ 230 MeV. 
\par
b) {\it {Neutron star mass and radius }}: 
The constraint also predicts a limiting neutron star mass of $M_{max}$ = 1.72 solar mass and a  radius, 
$R$ = 11 - 13 {\it {km}} for the `` canonical " neutron star. Recent observations of pulsar-white dwarf 
binaries at the Arecibo observatory suggest a pulsar mass for PSRJ0751+1807 of $M$ = 2.1$^{+0.4}_{-0.5}$ solar 
mass at a 95$\%$ confidence level \cite{NIC05}.
\par
c) {\it {Neutron star cooling }}: 
Furthermore, it predicts a direct URCA cooling for neutron stars above 1.4 times the solar mass. If such is
the case, then the enhanced cooling of a $M$ = 1.4 solar mass neutron star may provide strong evidence in favor 
of exotic matter in the core of a neutron star.
\par
These results have important implications for nuclear astrophysics and future experiments probing the 
properties of nuclei using beams of neutron-rich nuclei. The above constraint was obtained by studying the 
low density behavior of nuclear matter. Measurements at densities above normal nuclear matter should further 
constraint the form of the symmetry energy. Such measurements should yield consistent results when 
extrapolated to low densities.

\section{Summary and Conclusions}

In summary, a number of reactions were studied to investigate the density dependence of the symmetry 
energy in the equation of state of isospin asymmetric nuclear matter. The results were analyzed within the
framework of the dynamical and the statistical models of multifragmentation. It is observed that a dependence 
of the form C$_{sym}(\rho)$ = 31.6 ($\rho/\rho_{o})^{0.69}$ MeV, agrees reasonably with the experimental data indicating that a `` stiff " form of the symmetry 
energy provides a better description of the nuclear matter EOS at sub-nuclear densities. A comparison with several 
other independent studies shows that a constraint on the density dependence of the symmetry energy given 
as C$_{sym}(\rho)$ = C$_{sym}^{o}$($\rho/\rho_{o})^{\gamma}$, where C$_{sym}^{o}$ $\approx$ 31 - 33 MeV and 
$\gamma$ $\approx$ 0.55 - 0.69, can thereby be obtained. The present observation has important implications 
for astrophysics, as well as, nuclear physics studies to be carried out at future radioactive beam facilities 
worldwide. 

\section{acknowledgments}
This work was supported in part by the Robert A. Welch Foundation through grant No. A-1266, and the 
Department of Energy through grant No. DE-FG03-93ER40773. We also thank A. Botvina for the 
SMM code, and the Catania group for the BNV code.


\begin{thebibliography}{}
\bibitem{DIE03} A.E.L. Dieperink, Y. Dewulf, D. Van Neck, M. Waroquier, and V. Rodin, Phys. Rev. C {\bf 68}, 064307 (2003).
\bibitem{WIR88} R.B. Wiringa, V. Fiks and A. Fabrocini, Phys. Rev. C {\bf 38}, 1010 (1988).
\bibitem{LEE98} C.H. Lee, T.T.S. Kuo, G.Q. Li and G.E. Brown, Phys. Rev. C {\bf 57}, 3488 (1998).
\bibitem{LIU02} B. Liu, V. Greco, V. Baran, M. Colonna, and M. DiToro, Phys. Rev. C {\bf 65}, 045201 (2002).
\bibitem{KAI02} N. Kaiser, S. Fritsch and W. Weise, Nucl. Phys. A {\bf 697}, 255 (2002).
\bibitem{FUC06} C. Fuchs and H.H. Wolter, Eur. Phys. J. A {\bf 30}, 5 (2006).
\bibitem{BRO00} B.A. Brown, Phys. Rev. Lett {\bf 85}, 5296 (2000).
\bibitem{HOR01} C.J. Horowitz and J. Piekarewicz, Phys. Rev. Lett. {\bf 86}, 5647 (2001).
\bibitem{FUR02} R.J. Furnstahl, Nucl. Phys. A {\bf 706}, 85 (2002).
\bibitem{OYA98} K. Oyamatsu, I. Tanihata, Y. Sugahara, K. Sumiyoshi, and H. Toki, Nucl. Phys. A {\bf 634}, 3 (1998).
\bibitem{LAT91} J. Lattimer, C. Pethick, M. Prakash and P. Hansel, Phys. Rev. Lett. {\bf 66}, 2701 (1991).
\bibitem{LEE96} C. Lee, Phys. Rep. {\bf 275}, 255 (1996).
\bibitem{PET95} C.J. Pethick and D.G. Ravenhall, Annu. Rev. Nucl. Part. Sci. {\bf 45}, 429 (1995).
\bibitem{LAT00} J.M. Lattimer and M. Prakash, Phys. Rep. {\bf 333}, 121 (2000).
\bibitem{HIX03} W.R. Hix, O.E.B. Messer, A. Mezzacappa, M. Liebendorfer, J. Sampaio, K. Langanke, D.J. Dean, and G. Martinez-Pinedo, Phys. Rev. Lett. {\bf 91}, 201102 (2003).
\bibitem{LAT04} J.M. Lattimer and M. Prakash, Science, {\bf 304}, 536 (2004).
\bibitem{STE05} A.W. Steiner, M. Prakash, J.M. Lattimer and P.J. Ellis, Phys. Rep. {\bf 411}, 325 (2005).
\bibitem{HORO01} C.J. Horowitz, S.J. Pollock, P.A. Souder and R. Michaels, Phys. Rev. C {\bf 63}, 025501 (2001).
\bibitem{HOR02} C.J. Horowitz and J. Piekarewicz, Phys. Rev. C {\bf 66}, 55803 (2002).
\bibitem{STO03} J.R. Stone, J.C. Miller, R. Koncewicz, P.D. Stevenson and M.R. Strayer, Phys. Rev. C {\bf 68}, 034324 (2003).
\bibitem{LAT94} J. Lattimer {\it {et al.}}, Astrophys. J. {\bf 425}, 802 (1994).
\bibitem{SLA02} P. Slane, D.J. Helfand and S.S. Murray, Astrophys. J. L {\bf 571}, 45 (2002).
\bibitem{DAN02} P. Danielewicz, R. Lacey and W.G. Lynch, Science, {\bf 298}, 1592 (2002).
\bibitem{GSI}   GSI Conceptual Design Report, http://www.gsi.de/GSI-Future.
\bibitem{RIA}   RIA homepage, http://www.orau.org/ria.
\bibitem{BAL98} B.A. Li, C.M. Ko and W. Bauer, Int. J. Mod. Phys. E {\bf 7}, 147 (1998).
\bibitem{BAR05} V. Baran, M. Colonna, V. Greco and M. DiToro, Phys. Rep. {\bf 410}, 335 (2005).
\bibitem{TSANG01} M.B. Tsang, W.A. Friedman, C.K. Gelbke, W.G. Lynch, G. Verde, and H.S. Xu, Phys. Rev. Lett. {\bf 86}, 5023 (2001).
\bibitem{XU00}  H.S. Xu {\it {et al.}}, Phys. Rev. Lett. {\bf 85}, 716 (2000).
\bibitem{SHE03} D.V. Shetty, S.J. Yennello, E. Martin, A. Keksis, G.A. Souliotis, Phys. Rev. C {\bf 68}, 021602 (2003).
\bibitem{ONO03} A. Ono, P. Danielewicz, W.A. Friedman, W.G. Lynch and M.B. Tsang, Phys. Rev. C {\bf 68}, 051601 (2003).
\bibitem{QIN06} Q. Li, Z. Li, H. Stocker, Phys. Rev. C {\bf 73}, 051601 (2006).
\bibitem{LIU04} T.X. Liu {\it {et al.}}, Phys. Rev. C {\bf 69}, 014603 (2004).
\bibitem{COL05} M. Colonna and F. Matera, Phys. Rev. C {\bf 71}, 064605 (2005).
\bibitem{DOR06} C.O. Dorso, C.R. Escudero, M. Ison, and J.A. Lopez, Phys. Rev. C {\bf 73}, 044601 (2006).
\bibitem{BOT02} A.S. Botvina, O.V. Lozhkin and W. Trautmann, Phys. Rev. C {\bf 65}, 044610 (2002).
\bibitem{RAD02} Al.H. Raduta, Ad.R. Raduta, Phys. Rev. C {\bf 65}, 054610 (2002).
\bibitem{FRI90} W.A. Friedman, Phys. Rev. C {\bf 42}, 667 (1990).
\bibitem{GRO97} D.H.E. Gross, Phys. Rep. {\bf 279}, 119 (1997).
\bibitem{TSA01} M.B. Tsang {\it {et al.}}, Phys. Rev. C {\bf 64}, 054615 (2001).
\bibitem{TIA06} W.D. Tian {\it {et al.}}, arXiv: nucl-th/0601079 (2006).
\bibitem{ONO06} A. Ono, P. Danielewicz, W.A. Friedman, W.G. Lynch and M.B. Tsang, arXiv: nucl-ex/0507018 (2005).
\bibitem{ONO04} A. Ono, P. Danielewicz, W.A. Friedman {\it {et al.}}, Phys. Rev. C {\bf 70}, 041604 (2004).
\bibitem{JOH97} H. Johnston {\it {et al.}}, Phys. Rev. C {\bf 56}, 1972 (1997).
\bibitem{YEN94} S.J. Yennello {\it {et al.}}, Phys. Lett. B {\bf 321}, 15 (1994).
\bibitem{SCH95} R.P. Schmitt {\it {et al.}}, Nucl. Instrum. Methods Phys. Res. A  {\bf 354}, 487 (1995).
\bibitem{OGI89} C.A. Ogilvie, D.A. Cebra, J. Clayton, S. Howden, J. Karn, A. Vander Molen, G.D. Westfall, W.K. Wilson, and J.S. Winfield, Phys. Rev. C {\bf 40}, 654 (1989).
\bibitem{YENN94} S.J. Yennello {\it {et al.}}, {\it {Proc. of the 10th Winter Workshop on Nuclear Dynamics, Snowbird, UT,}} edited by W. Bauer (1994).
\bibitem{GER04} E. Geraci {\it {et al.}}, Nucl. Phys. A {\bf 732}, 173 (2004).
\bibitem{CHE05} L.W. Chen, C.M. Ko and B.A. Li, Phys. Rev. Lett. {\bf 94}, 032701 (2005).
\bibitem{BON95} J.P. Bondorf {\it {et al.}}, Phys. Rep. {\bf 257}, 133 (1995). 
\bibitem{BOT95} A.S. Botvina {\it {et al.}}, Nucl. Phys. {\bf A 584}, 737 (1995).
\bibitem{BOT87} A.S. Botvina, A.S. Iljinov, I.N. Mishustin, J.P. Bondorf, R. Donangelo, and K. Sneppen, Nucl. Phys. {\bf A 475}, 663 (1987).
\bibitem{BAR02} V. Baran, M. Colonna, M. Di~Toro, V. Greco, M. Zielinska-Pfabe and H.H.~Wolter, Nucl. Phys. A {\bf 703}, 603 (2002).
\bibitem{NAT02} J.B. Natowitz, R. Wada, K. Hagel, T. Keutgen, M. Murray, A. Makeev, L. Qin, P. Smith, and C. Hamilton, Phys. Rev. C {\bf 65}, 034618 (2002); {\it {and references therein}}. 
\bibitem{BOT01} A.S. Botvina and I.N. Mishustin, Phys. Rev. C {\bf 63}, 061601 (2001).
\bibitem{BON85} J.P. Bondorf, R. Donangelo, I.N. Mishustin and H. Schultz, Nucl. Phys. A {\bf 444}, 460 (1985). 
\bibitem{BON98} J.P. Bondorf, A.S. Botvina and I.N. Mishustin, Phys. Rev. C {\bf 58}, 27 (1998). 
\bibitem{TAN01} W.P. Tan {\it {et al.}}, Phys. Rev. C {\bf 64}, 051901 (2001).
\bibitem{IGL06} J. Iglio, D.V. Shetty, S.J. Yennello, G.A. Souliotis, M. Jandel, A. Keksis, S. Soisson, B. Stein and S. Wuenschel, Phys. Rev. C {\bf 74}, 024605 (2006).
\bibitem{SHE05} D.V. Shetty, A.S. Botvina, S.J. Yennello, A. Keksis, E. Martin and G.A. Souliotis, Phys. Rev. C {\bf 71}, 024602 (2005).
\bibitem{FEV05} A. LeFevre {\it {et al.}}, Phys. Rev. Lett. {\bf 94}, 162701 (2005).
\bibitem{HEN05} D. Henzlova, A.S. Botvina, K.H. Schmidt, V. Henzl, P. Napolitani, and M.V. Ricciardi, arXiv: nucl-ex/0507003 (2005). 
\bibitem{BUY05} N. Buyukcizmeci, R. Ogul, and A.S. Botvina, Eur. Phys. J. {\bf 25}, (2005) 57. 
\bibitem{SOB04} L.G. Sobotka, R.J. Charity, J. Toke and W.U. Schroder, Phys. Rev. Lett. {\bf 93}, 132702 (2004).
\bibitem{HAS86} R. Hasse and P. Schuck, Phys. Lett. B {\bf 179}, 313 (1986).
\bibitem{SHL90} S. Shlomo and J.B. Natowitz, Phys. Lett. B {\bf 252}, 187 (1990).
\bibitem{SHL91} S. Shlomo and J.B. Natowitz, Phys. Rev. C {\bf 44}, 2878 (1991).
\bibitem{NATO02} J.B. Natowitz, K. Hagel, Y. Ma, M. Murray, L. Qin, S. Shlomo, R. Wada, and J. Wang, Phys. Rev. C {\bf 66}, 031601 (2002). 
\bibitem{VIO04} V.E. Viola, K. Kwiatkowski, J.B. Natowitz, and S.J. Yennello, Phys. Rev. Lett. {\bf 93}, 132701 (2004).
\bibitem{BAL01} {\it {Isospin Physics in Heavy Ion Collisions at Intermediate Energies,}} edited by B.-A. Li and W. Schroder (Nova Science, New York, 2001).
\bibitem{BAL06} B.A. Li and L.W. Chen, Phys. Rev. C {\bf 74}, 034610 (2006).
\bibitem{SHE04} D.V. Shetty, S.J. Yennello, A.S. Botvina, G.A. Souliotis, M. Jandel, E. Bell, A. Keksis, S. Soisson, B. Stein, and J. Iglio , Phys. Rev. C {\bf 70}, 011601 (2004).
\bibitem{SHET05} D.V. Shetty, S.J. Yennello, and G.A. Souliotis, Phys. Rev. C {\bf 75}, 034602 (2007); arXiv: nucl-ex/0505011 (2005).
\bibitem{SHET06} D.V. Shetty, S.J. Yennello, G.A. Souliotis, A.L. Keksis, S.N. Soisson, B.C. Stein, and S. Wuenschel, Phys. Rev. C (2007) (Submitted); arXiV: nucl-ex/0606032 (2006).
\bibitem{TOD05} B.G. Todd-Rutel and J. Piekarewicz, Phys. Rev. Lett {\bf 95}, 122501 (2005).
\bibitem{PIE05} J. Piekarewicz (Private Communication).
\bibitem{PIE06} J. Piekarewicz, {\it {Proc. of the International Conference on Current Problems in Nuclear Physics and Atomic Energy, Kyiv, Ukraine}}, (May 29 - June 3, 2006).
\bibitem{TSA04} M.B. Tsang {\it {et al.,}} Phys. Rev. Lett. {\bf 92}, 062701 (2004).
\bibitem{BAL05} B.A. Li and L.W. Chen, Phys. Rev. C {\bf 72}, 064611 (2005).
\bibitem{DAN04} P. Danielewicz, arXiv : nucl-th/0411115 (2004).
\bibitem{HEI00} H. Heiselberg and M. Hjorth-Jensen, Phys. Rep. {\bf 328}, 237 (2000).
\bibitem{AKM97} A. Akmal and V.R. Pandharipande, Phys. Rev. C {\bf 56}, 2261 (1997).
\bibitem{AKM98} A. Akmal, V.R. Pandharipande and D.G. Ravenhall, Phys. Rev. C {\bf 58}, 1804 (1998).
\bibitem{FUCH06} C. Fuchs (Private Communication); E.N.E. van~Dalen, C. Fuchs, and A. Faessler, Nucl. Phys. A {\bf 744}, 227 (2004).
\bibitem{KHO05} D.T. Khoa and H.S. Than, Phys. Rev. C {\bf 71}, 044601 (2005).
\bibitem{FAM06} M.A. Famiano {\it {et al.,}} Phys. Rev. Lett. {\bf 97}, 052701 (2006).
\bibitem{BAL97} B.A. Li, C. Ko, and Z. Ren, Phys. Rev. Lett. {\bf 78}, 1644 (1997).
\bibitem{BALI06} B.A. Li, L.W. Chen, G.C. Yong, and W. Zuo, Phys. Lett. B {\bf 634}, 378 (2006).
\bibitem{HUD03} S. Hudan {\it {et al.}}, Phys. Rev. C {\bf 67}, 064613 (2003).
\bibitem{COL06} M. Colonna and M.B. Tsang, Eur. Phys. J. A {\bf 30}, 165 (2006).
\bibitem{DAN03} P. Danielewicz, Nucl. Phys. A{\bf 727}, 233 (2003).
\bibitem{ZUO99} W. Zuo, I. Bombaci, and U. Lombardo, Phys. Rev. C {\bf 60}, 024605 (1999).
\bibitem{BRA85} M. Brack, C. Guet, and H.B. Hakansson, Phys. Rep. {\bf 123}, 276 (1985).
\bibitem{PEA00} J.M. Pearson and R.C. Nayak, Nucl. Phys. A {\bf 668}, 163 (2000).
\bibitem{MYE66} W.D. Myers and W.J. Swiatecki, Nucl. Phys. {\bf 81}, 1 (1966).
\bibitem{POM03} K. Pomorski and J. Dudek, Phys. Rev. C {\bf 67}, 044316 (2003).
\bibitem{HOR06} C.J. Horowitz, Eur. Phys. J. A {\bf 30}, 303 (2006).
\bibitem{NIC05} D.J. Nice {\it {et al.,}} Astrophys. J. {\bf 634}, 1242 (2005).
\end{thebibliography}
\end{document}